\documentclass[epj]{svjour}

\usepackage{epsfig,graphics,graphicx}
\usepackage{clshan-math,clshan-font,clshan-dm-dd}
\def\lsim{\:\raisebox{-0.5ex}{$\stackrel{\textstyle<}{\sim}$}\:}
\def\gsim{\:\raisebox{-0.5ex}{$\stackrel{\textstyle>}{\sim}$}\:}
\usepackage{fancyhdr}
\def\makeheadbox{{
 }}
\def\mytitle{My title} 
\def\myauthors{My name}  
\def\mytype{My type of session}
\def\mysession{My session}

\def\mytitle{Determining the WIMP Mass from Direct DM Detection Data}
\def\myauthors{C.~L.~Shan and M.~Drees}
\def\mytype{Contributed Talk}    
\def\mysession{Cosmology and Astrophysics}

\begin{document}
\title{Determining the WIMP Mass from Direct Dark Matter Detection Data}
\author{Chung-Lin Shan\inst{1}%
\thanks{\emph{Email:} cshan@th.physik.uni-bonn.de}%
 \and
 Manuel Drees\inst{1}
\thanks{\emph{Email:} drees@th.physik.uni-bonn.de}%
}                     
%
%
\institute{Physikalisches Inst. der Univ. Bonn, Nussallee 12, 53115 Bonn, Germany}
%
\date{\today}
\abstract{
 Weakly interacting massive particles (WIMPs)
 are one of the leading candidates for Dark Matter.
 So far we can use direct Dark Matter detection
 to estimate the mass of halo WIMPs
 only by fitting predicted recoil spectra to future experimental data.
 Here we develop a model--independent method for determining the WIMP mass
 by using experimental data directly.
 This method is independent of the as yet unknown WIMP density near the Earth
 as well as of the WIMP--nuclear cross section
 and can be used to extract information about WIMP mass with ${\cal O}(50)$ events.
\PACS{
      {95.35.+d}{Dark Matter}   \and
      {29.85.Fj}{data analysis}
     } 
} 
\maketitle
%

\section{Introduction}
\label{intro}
%
%
%
 Today astrophysicists have strong evidence \cite{evida}-\cite{bullet} to believe that
 a large fraction (more than 80\%) of the matter in the Universe is dark
 (i.e., interacts at most very weakly with electromagnetic radiation).
 The dominant component of this cosmological Dark Matter (DM) must be
 due to some yet to be discovered, non--baryonic particles.
 Weakly interacting massive particles (WIMPs) $\chi$
 with masses between 10 GeV and a few TeV
 are one of the leading candidates for DM \cite{susydm}.

 Currently,
 the most promising method to detect different WIMP candidates
 is the direct detection of the recoil energy deposited in a low--background laboratory detector
 by elastic scattering of ambient WIMPs off the target nuclei \cite{detaa}, \cite{detab}.
 The differential rate for elastic WIMP--nucleus scattering is given by \cite{susydm}:
\beq \label{eqn2101}
   \dRdQ
 = \calA \FQ \intvmin \bfrac{f_1(v)}{v} dv\, .
\eeq
 Here $R$ is the direct detection event rate,
 i.e., the number of events per unit time and unit mass of detector material,
 $Q$ is the energy deposited in the detector,
 $F(Q)$ is the elastic nuclear form factor,
 and $f_1(v)$ is the one--dimensional velocity distribution function of the WIMPs
 impinging on the detector.
 The constant coefficient $\calA$ is defined as
\beq \label{eqn2102}
        \calA
 \equiv \frac{\rho_0 \sigma_0}{2 \mchi m_{\rm r,N}^2}\, ,
\eeq
 where $\rho_0$ is the WIMP density near the Earth
 and $\sigma_0$ is the total cross section ignoring the form factor suppression.
 The reduced mass $m_{\rm r,N}$ is defined by
\beq \label{eqn2103}
        m_{\rm r,N}
 \equiv \frac{\mchi \mN}{\mchi+\mN}\, ,
\eeq
 where $\mchi$ is the WIMP mass and $\mN$ that of the target nucleus.
 Finally,
 $\vmin$ is the minimal incoming velocity of incident WIMPs
 that can deposit the energy $Q$ in the detector:
\beq \label{eqn2104}
   \vmin
 = \alpha \sqrt{Q}\, ,
\eeq
 where we define
\beq \label{eqn2105}
        \alpha
 \equiv \sfrac{\mN}{2 m_{\rm r,N}^2}\, .
\eeq

 So far most theoretical analyses of direct WIMP detection
 have predicted the detection rate for a given (class of) WIMP(s),
 based on a specific model of the galactic halo.
 This can be used to estimate the mass of halo WIMPs
 only by fitting the predicted recoil spectra to future experimental data,
 e.g., \cite{Green07}.
 The goal of our work is to develop methods
 which allow to extract information on halo WIMPs
 by using the experimental data directly.
 In our earlier work \cite{DMDD}
 we used a time--averaged recoil spectrum,
 assumed that no directional information exists,
 and derived an expression for estimating moments of
 the normalized one--dimensional velocity distribution function of halo WIMPs:
\beqn \label{eqn2208}
 \expv{v^n}
 \eqnequiv \int_{v_{\rm min}(\Qthre)}^\infty v^n f_1(v) \~ dv
    \non\\
 \= \alpha^n
    \bfrac{2 \Qthre^{(n+1)/2} \rthre+(n+1) I_n \FQthre}{2 \Qthre^{1/2} \rthre+I_0 \FQthre}\, ,
    \non\\
\eeqn
 where $Q_{\rm thre}$ is the threshold energy of the detector,
 $\rthre \equiv (dR/dQ)_{Q = \Qthre}$
 is an estimated value of the scattering spectrum at $Q = \Qthre$.
 $I_n$ can be either determined
 from a given expression (e.g., a fit to data) for the recoil spectrum:
\beq \label{eqn2110}
   I_n
 = \int_{\Qthre}^{\infty} \frac{Q^{(n-1)/2}}{\FQ} \adRdQ \~ dQ\, ,
\eeq
 or estimated directly from the measured recoil energy:
\beq \label{eqn2210}
   I_n
 = \sum_a \frac{Q_a^{(n-1)/2}}{F^2(Q_a)}\, ,
\eeq
 where the sum runs over all events in the data set.
 Note that
 all these expressions are independent of the as yet unknown WIMP density near the Earth
 as well as of the WIMP--nucleus cross section.
 (More details about
  the reconstruction of the velocity distribution function of halo WIMPs
  and the estimate of its moments
  as well as all formulae needed can be found in Ref.~\cite{DMDD}.)
\section{Determining the WIMP mass}
\label{secmchi}
%
 In this section we present a method for determining the WIMP mass
 based on the estimate of the moments $\expv{v^n}$
 from two (or more) experimental data sets with different target materials.
 The basic idea is that,
 from independent direct WIMP detection experiments with different target nuclei,
 the measured recoil spectra should lead to
 the same (moments of the) velocity distribution function of incident WIMPs,
 when the threshold energies of these experiments are low enough.
 Note that this can be done {\em independent of the detailed particle physics model},
 which determines the value of $\sigma_0$ for the target nuclei.
 However,
 one will need to know the form factor of each target nucleus,
 which strongly depends on whether spin--dependent or spin--independent interaction dominates.

 Moreover,
 as shown in Ref.~\cite{DMDD},
 because data binning is not required for estimate of the moments $\expv{v^n}$,
 some non--trivial information can already be extracted from ${\cal O}(20)$ events,
 it thus seems reasonable to expect that
 one can also obtain meaningful information about the WIMP mass with pretty few events.
\subsection{Neglecting threshold energies}
\label{secmchia}
 For the case that the threshold energy $\Qthre$ can be neglected,
 the $n-$th moment of the velocity distribution function, $\expv{v^n}$,
 in Eq.(\ref{eqn2208}) can be reduced to
\beq \label{eqn3101}
   \expv{v^n}
 = \alpha_X^n \bfrac{(n+1) \InX}{\IzX}
 = \alpha_Y^n \bfrac{(n+1) \InY}{\IzY},
\eeq
 where $X$ and $Y$ are two target nuclei,
 $\InX$, $\IzX$ and so on can be estimated by Eq.(\ref{eqn2110}) or (\ref{eqn2210}).
 Note that
 the form factor $\FQ$ in Eq.(\ref{eqn2110}) or (\ref{eqn2210})
 for estimating $\InX$ and $\InY$ are different.
 According to the definition of $\alpha$ in Eq.(\ref{eqn2105})
 with the expression of the reduced mass $m_{\rm r,N}$ in Eq.(\ref{eqn2103}),
 and using some simple algebra,
 one can solve the WIMP mass as
\beq \label{eqn3103}
   \mchi
 = \frac{\sqrt{\mX \mY}-\mX \calRn}{\calRn-\sqrt{\mX/\mY}}\, ,
\eeq
 where we have defined
\beq \label{eqn3104}
        \calRn
 \equiv \frac{\alpha_Y}{\alpha_X}
 =      \abrac{\frac{\InX}{\IzX} \cdot \frac{\IzY}{\InY}}^{1/n},
       ~~~~~
        n
 \ne    0,~-1.
\eeq

\begin{figure}[t]
\begin{center}
\includegraphics[width=0.5\textwidth]{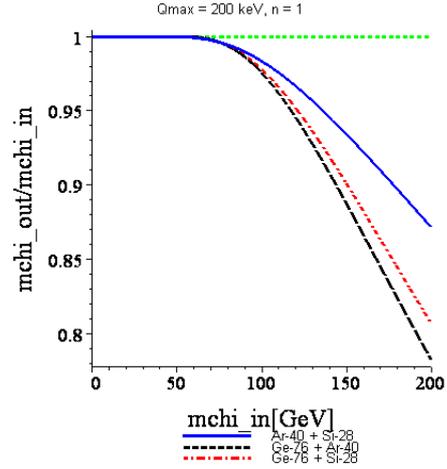}
\end{center}
\caption{
 The curves show
 the ratios of the reproduced WIMP masses estimated by Eq.(\ref{eqn3103})
 with different combinations of target nuclei
 to the input (true) one as functions of the input WIMP mass.
 The theoretical predicted recoil spectrum
 for the shifted Maxwellian velocity distribution function \cite{susydm}, \cite{DMDD}
 with the Woods--Saxon form factor \cite{FQb}, \cite{susydm} has been used
 ($\mN = 70.6~{\rm GeV}/c^2$ for $\rmXA{Ge}{76}$,
  $v_0 = 220$ km/s,
  $v_e = 231$ km/s).
 The solid (blue) line,
 the dashed (black) line,
 and the dash--dotted (red) line
 are for $\rmXA{Ar}{40}$ + $\rmXA{Si}{28}$,
 $\rmXA{Ge}{76}$ + $\rmXA{Ar}{40}$,
 and $\rmXA{Ge}{76}$ + $\rmXA{Si}{28}$ combination,
 respectively.} 
\label{fig3101}
\end{figure}
 Fig.~\ref{fig3101} shows the ratios of the reproduced WIMP masses
 estimated by Eq.(\ref{eqn3103}) with different combinations of target nuclei
 to the input (true) one as functions of the input WIMP mass.
 $\rmXA{Si}{28}$, $\rmXA{Ar}{40}$, and $\rmXA{Ge}{76}$ have been chosen
 as three target nuclei and
 thus three combinations for $\calRn$ defined in Eq.(\ref{eqn3104})
 with $n = 1$ are shown.
 $\calRn$ has been estimated by the integral form of $I_n$ in Eq.(\ref{eqn2110})
 with a maximal measuring energy of 200 keV.
 In Fig.~\ref{fig3101}
 we can see obviously a deviation of the reproduced WIMP mass
 from the input (true) one
 as input $\mchi \gsim 60~{\rm GeV}/c^2$.
 The heavier the nuclear masses of two target nuclei,
 the larger the deviation from the true WIMP mass.
 This is caused by introducing the maximal measuring energy for estimating $I_n$.
 For $n = 1$ and input $\mchi = 200~{\rm GeV}/c^2$,
 the deviation with $Q_{\rm max} = 200$ keV is around 20\%.
 However,
 for input $\mchi \lsim 120~{\rm GeV}/c^2$,
 this deviation will be less than 5\%,
 and if $Q_{\rm max} = 250$ keV or 300 keV,
 this deviation will be reduced to 10\% or even only 5\%.

 Furthermore,
 the statistical error on the reproduced WIMP mass
 can be obtained from Eq.(\ref{eqn3103}) directly as
\beqn \label{eqn3105}
        \sigma(\mchi)
 \=     \frac{\calRn \sqrt{\mX/\mY} \vbrac{\mX-\mY}}{|n| \abrac{\calRn-\sqrt{\mX/\mY}}^2}
        \non\\
 \conti ~~ \times
        \bBiggl{ \frac{\sigma^2(\InX)}{\InX^2}
                +\frac{\sigma^2(\IzX)}{\IzX^2}
                -\frac{2 {\rm cov}(\IzX,\InX)}{\IzX \InX}}
        \non\\
 \conti ~~~~~~~~~~~~ 
        \bbiggr{+(X \lto Y)}^{1/2}.
\eeqn
 The formulae for estimating
 $\sigma^2(\InX) = {\rm cov}(\InX,\InX)$, ${\rm cov}(\IzX,\InX)$ and so on
 can be found in Ref.~\cite{DMDD}.

\begin{figure}[t]
\begin{center}
\includegraphics[width=0.5\textwidth]{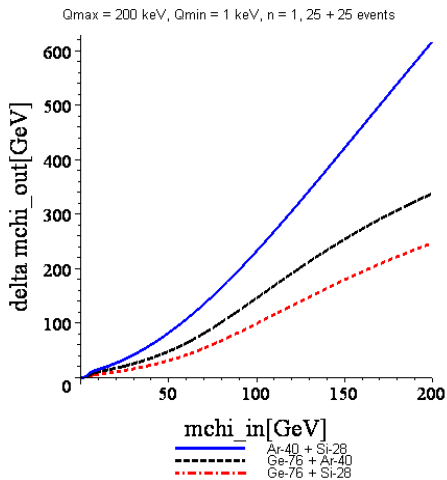}
\end{center}
\caption{
 The curves show the statistical errors estimated by Eq.(\ref{eqn3105})
 with different combinations of target nuclei
 as functions of the input WIMP mass.
 Each experiment has 25 events, i.e., totally 50 events.
 Parameters and indications of the lines as in Fig.~\ref{fig3101}.} 
\label{fig3102}
\end{figure}
 Fig.~\ref{fig3102} shows the 1$-\sigma$ statistical errors estimated by Eq.(\ref{eqn3105})
 with three different combinations of target nuclei
 as functions of the input (true) WIMP mass.
 Each experiment has 25 events, i.e., totally 50 events.
 Note that,
 in order to estimate ${\rm cov}(I_n,I_m)$ above,
 a threshold energy $Q_{\rm min} = 1$ keV for both nuclei has been given.
 In Fig.~\ref{fig3102}
 we can observe that
 {\em the larger the mass difference between two detector nuclei,
 the smaller the statistical error will be}.
 Hence,
 the combination with the largest mass difference,
 $\rmXA{Ge}{76}$ + $\rmXA{Si}{28}$,
 will have the smallest statistical error.
 On the other hand,
 despite of the factor $1/|n|$ in Eq.(\ref{eqn3105}),
 it has been found that
 the statistical errors increase generally with increasing $n$.
 Hence,
 $n = 1$ should be the best choice for $\mchi$ and $\sigma(\mchi)$
 in Eqs.(\ref{eqn3103}) and (\ref{eqn3105}),
 respectively.

\begin{figure}[t]
\begin{center}
\includegraphics[width=0.5\textwidth]{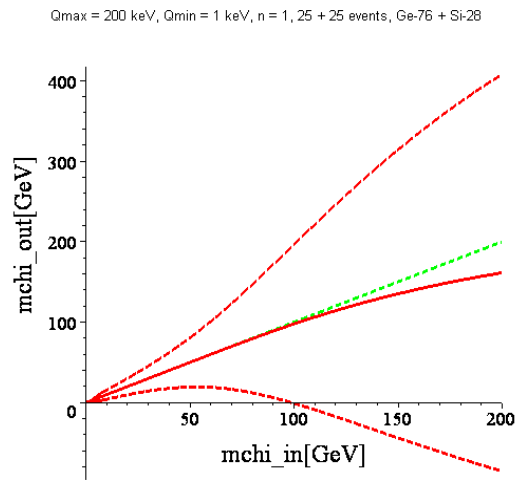} \\ \vspace{0.8cm}
\includegraphics[width=0.5\textwidth]{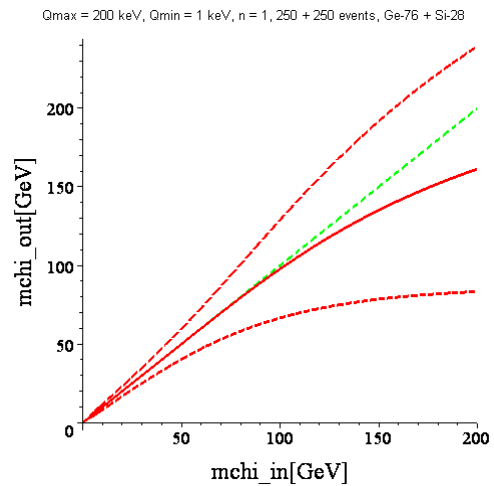}
\end{center}
\caption{
 The reproduced WIMP mass with the statistical error
 by using $\rmXA{Ge}{76}$ and $\rmXA{Si}{28}$ as two target nuclei
 as a function of the input WIMP mass.
 The solid (red) line indicates the reproduced WIMP mass estimated by Eq.(\ref{eqn3103}),
 the dashed (red) lines indicate the 1$-\sigma$ statistical error
 estimated by Eq.(\ref{eqn3105}).
 The straight dash--dotted (green) line indicates the input (true) WIMP mass.
 Each experiment has 25 (250) events,
 i.e., totally 50 (500) events,
 in the upper (lower) frame.
 Parameters as in Fig.~\ref{fig3101}.}
\label{fig3103}
\end{figure}
 Figs.~\ref{fig3103} show the reproduced WIMP mass with the 1$-\sigma$ statistical error
 by using $\rmXA{Ge}{76}$ and $\rmXA{Si}{28}$ as two target nuclei
 as a function of the input (true) WIMP mass.
 From the upper frame,
 it can be found that,
 despite of the very few (25 + 25, totally 50) events
 and correspondingly very large statistical error,
 for $\mchi \le 100~{\rm GeV}/c^2$,
 one can already extract some meaningful information on the WIMP mass.
 For example,
 for $\mchi = 25~{\rm GeV}/c^2$ and $\mchi = 50~{\rm GeV}/c^2$,
 we will reproduce $\mchi \simeq (25 \pm 13)~{\rm GeV}/c^2$
 and $\mchi \simeq (50 \pm 31)~{\rm GeV}/c^2$.
 For the case with 500 (250 + 250) total events,
 the statistical error will be reduced to less than 5 and 10 ${\rm GeV}/c^2$,
 respectively!
 Certainly,
 as shown in the lower frame of Figs.~\ref{fig3103},
 for the case with 500 total events,
 the deviation of the reproduced WIMP mass from the input one becomes important.
 Nevertheless,
 in practice,
 an experiment with more than 200 events
 should have a larger maximal measuring energy,
 and,
 as discussed above,
 the deviation can (should) be strongly reduced.

 For the simplified simulations with the integral form of $I_n$ presented above,
 the event numbers from both experiments have been considered to be equal.
 Practically
 experiments with the higher mass nuclei, e.g., Ge or Xe,
 are expected to measured (much) more signal events.
 However,
 some detailed simulations show that,
 the reproduced WIMP mass with the statistical uncertainty
 estimated by two experiments
 with different maximal measuring energies
 and different exposures, or, equivalently, different event numbers,
 will be modified only slightly from that shown in Figs.~\ref{fig3103}.
%
 Moreover,
 a detailed analysis of contributions from different terms of $\sigma(\mchi)$ show that
 one {\em can not} reduce the statistical error of $\mchi$ estimated by Eq.(\ref{eqn3105})
 by {\em improving only one experiment} with even very large event number,
 since the contribution from the other (poor) experiment will dominate the error.
\subsection{Taking into account threshold energies}
\label{secmchib}
 For the case that $\Qthre$ in Eq.(\ref{eqn2208}) can not be neglected,
 $\calRn$ defined in Eq.(\ref{eqn3104})
 should be modified to the following general form:
\beqn \label{eqn3201}
 \calRn
 \=     \bfrac{2 \QthreX^{(n+1)/2} \rthreX+(n+1) \InX \FQthreX}
              {2 \QthreX^{    1/2} \rthreX+      \IzX \FQthreX}^{1/n}
        \non\\
        \non\\
 \conti ~~~~~~~~ \times 
        (X \lto Y)^{-1}\, ,
\eeqn
 where $n \ne 0$.
 In this general form of $\calRn$
 there are totally six variables:
 $\rthreX$, $\InX$, $\IzX$ and the other three for nucleus $Y$.
 This should generally produce a larger statistical error
 than that estimated by Eq.(\ref{eqn3105})
 due to the contribution from $\rthre$.
 However,
 one can practically reduce the number of variables by choosing $n = -1$:
\beq \label{eqn3202}
   \calRma
 = \frac{\rthreY}{\rthreX}
   \bfrac{2 \QthreX^{1/2} \rthreX+ \IzX \FQthreX}{2 \QthreY^{1/2} \rthreY+ \IzY \FQthreY}\, .
\eeq
%
%
\section{Summary and Conclusions}
 In this paper
 we have presented a method
 which allows to extract information on the WIMP mass
 from the recoil energy measured in elastic WIMP--nucleus scattering experiments directly.
 In the long term
 this information can be used to
 constrain e.g., SUSY models in the elementary particle physics
 and compare with information from future collider experiments.

 Our method for determining the WIMP mass
 by combining two (or more) experiments with different detector materials
 is independent of the as yet unknown WIMP density near the Earth
 as well as of the WIMP--nucleus cross section.
 The only information which one needs is the measured recoil energy.

 Due to the maximal measuring energy of the detector,
 there will be a deviation of the reproduced WIMP mass from the true one.
 Nevertheless,
 for experiments with very few events and thus a quite large statistical error
 in the near future,
 this deviation should not affect the reproduced WIMP mass very significantly.
 Moreover,
 the numerical analysis shows also that,
 for WIMP masses $\le 100~{\rm GeV}/c^2$
 some meaningful information on the WIMP mass can already be extracted
 from ${\cal O}(50)$ total (each experiment ${\cal O}(25)$) events.

 The analyses of this work are based on several simplified assumptions.
 First,
 all experimental systematic uncertainties,
 as well as the uncertainty on the measurement of the recoil energy
 have been ignored.
 Comparing with large statistical uncertainty
 this should be a quite good approximation.
 Second,
 the analysis treats each recorded event as signal,
 i.e., background has been ignored altogether.
 This may in fact not be unrealistic for modern detectors.
 Third,
 each detector consists of a single isotope.
 This is quite realistic for the 
 semiconductor 
 detectors.
 For detectors containing more than one nucleus,
 by simultaneously measuring two signals,
 one might be able to tell on an event--by--event basis
 which kind of nucleus has been struck.

%
%
\subsubsection*{Acknowledgments}
 This work
 was partially supported by the Marie Curie Training Research Network ``UniverseNet''
 under contract no. MRTN-CT-2006-035863,
 as well as by the European Network of Theoretical Astroparticle Physics ENTApP ILIAS/N6
 under contract no. RII3-CT-2004-506222.
%
%

%

%
%

\end{document}